\documentclass[
  aps, prd, twocolumn, superscriptaddress, nofootinbib,
  amsmath, amssymb, floatfix
]{revtex4-2}

\usepackage{graphicx}
\usepackage[colorlinks=true,linkcolor=blue,citecolor=blue,urlcolor=blue]{hyperref}

\newcommand{\dd}{\mathrm{d}}
\newcommand{\jump}[1]{\left[#1\right]^{+}_{-}}
\newcommand{\half}{\tfrac{1}{2}}

\begin{document}

\title{Exact radial stability threshold for a Minkowski--Schwarzschild thin shell}

\author{Anuar Idrissov}
\email{anuar.idrissov@gmail.com}
\affiliation{Instituto de Ciencias Nucleares, Universidad Nacional Aut\'onoma de M\'exico, Mexico}
\affiliation{Fesenkov Astrophysical Institute, Observatory 23, 050020, Almaty, Kazakhstan}
\affiliation{Al-Farabi Kazakh National University, Al-Farabi Ave.\ 71, 050040, Almaty, Kazakhstan}
\date{\today}

\begin{abstract}
We study the stability of a thin shell of ordinary matter that separates a flat
interior from a Schwarzschild exterior, which is the simplest model of a
subluminal warp drive. The matter content of the shell is fixed by the junction
conditions and satisfies all the standard energy conditions except very close to
the horizon. For a shell made of causal matter we find an exact radius
below which it cannot be stable under radial perturbations. We then place a black
hole inside the shell and show that this radius grows with the black-hole mass,
so that the flat interior is the most favourable case. Since such shells are
known to be unstable to non-radial perturbations, our result marks the boundary
of the region in which any stabilising mechanism would have to operate.
\end{abstract}

\maketitle

%% ===================================================================
\section{Introduction}
\label{sec:intro}
%% ===================================================================

Warp drives are among the most studied exotic solutions of general relativity.
Since the original proposal of Alcubierre~\cite{Alcubierre1994} and the later
variants of Nat\'ario~\cite{Natario2002} and Van Den Broeck~\cite{VanDenBroeck1999},
it has been known that these spacetimes need matter that violates the null energy
condition, and it was later shown that this is a generic feature of the
construction~\cite{SantiagoSchusterVisser2022}. In all these models one first
prescribes the metric and then infer the energy-momentum tensor required to
support it, with no guarantee that such matter exists or that the configuration
survives small perturbations.

A different point of view was put forward by Bobrick and
Martire~\cite{BobrickMartire2021}, who noticed that once superluminal motion is
given up, a warp drive is essentially a shell of matter enclosing a region of flat
space that moves inertially. Along these lines, Fuchs \emph{et al.}~\cite{Fuchs2024}
constructed numerically, with the tools of Ref.~\cite{Helmerich2024}, a subluminal
warp solution that satisfies all the energy conditions (see also
Ref.~\cite{SchusterSantiagoVisser2023} for the ADM mass of warp spacetimes). In
this picture the exotic-matter problem disappears and two simpler questions
remain: what kind of matter can support the shell, and is the shell stable?

The simplest model that answers both questions exactly is a thin shell joining a
Minkowski interior to a Schwarzschild exterior. In the thin-shell limit the
matter on the wall is not postulated but determined by the Darmois--Israel
junction conditions~\cite{Israel1966,MusgraveLake1996,Poisson2004}, and the radial
motion reduces to a one-dimensional problem with an effective
potential~\cite{Visser1989,PoissonVisser1995}. General stability criteria for such
junctions were given in Refs.~\cite{IshakLake2002,LoboCrawford2005}. Shells
surrounding a black hole of mass $m$ were studied by Brady, Louko and
Poisson~\cite{BradyLoukoPoisson1991}, who found that a stable shell made of causal
matter always lies outside the circular photon orbit, while the static solution
for vanishing interior mass was obtained in Ref.~\cite{FHK1990}. More recently,
junctions between two Schwarzschild regions have been surveyed numerically as a
function of the mass ratio~\cite{RippentropBeraIshak2025}, and the family has
been extended to include a cosmological constant~\cite{Efstratiou2026}. The flat
interior is the $m\to0$ end of this family.

Radial stability, however, is only a necessary condition. Yang, Bonga and
Pan~\cite{YangBongaPan2023} showed that membranes with positive density and
pressure are unstable to perturbations of high angular number, and Pitre,
Schneider and Poisson~\cite{PSP2026} recently computed the quasinormal modes of
exactly the configuration considered here, finding a growing mode for all sampled
parameters and all multipoles $\ell\ge2$. The radial sector is nevertheless the
only one that can be solved exactly, and it tells us where in parameter space a
stabilising mechanism would have to act.

Our aim in the present work is to obtain analytically the radial stability threshold of the
Minkowski--Schwarzschild shell and to understand how it fits into
the family of shells around a black hole. We show that the critical sound speed
of the wall equals the slope of the sequence of static solutions, which gives a
short derivation valid for any pair of bulk metrics. For causal wall material the
threshold reduces to a cubic equation, whose root gives $a_{\rm crit}\simeq2.37\,M$.
We then prove that, with a black hole of mass $m$ inside, the threshold grows
monotonically with $m$ and always lies between $3m$ and $3M$, which reconciles
our result with the photon-orbit bound of Ref.~\cite{BradyLoukoPoisson1991}.

The paper is organized as follows. In Sec.~\ref{sec:setup} we describe the model
and the sense in which it is a warp drive. In Sec.~\ref{sec:junction} we derive
the surface stresses and discuss the energy conditions. Section~\ref{sec:dynamics}
is devoted to radial stability, and Sec.~\ref{sec:interp} to the family with a
black hole in the interior. Section~\ref{sec:discussion} contains a discussion of
the results. Throughout the paper we use units $G=c=1$ and signature
$(-,+,+,+)$.

%% ===================================================================
\section{The model}
\label{sec:setup}
%% ===================================================================

We consider a timelike hypersurface $\Sigma$ separating an interior region
$\mathcal{V}^{-}$ from an exterior region $\mathcal{V}^{+}$, with
\begin{equation}
  \dd s^{2}_{\pm} = -f_{\pm}(r)\,\dd t_{\pm}^{2}
   + \frac{\dd r^{2}}{f_{\pm}(r)} + r^{2}\dd\Omega^{2},
  \label{eq:bulk}
\end{equation}
where
\begin{equation}
  f_{-}(r)=1,\qquad f_{+}(r)=1-\frac{2M}{r}.
  \label{eq:fpm}
\end{equation}
Here $r$ is the areal radius on both sides, so the first junction condition is
satisfied by matching $r$ across $\Sigma$. The interior is a ball of flat space
and the exterior is Schwarzschild with ADM mass $M$, so that all the matter is in
the wall. We take the shell radius $a>2M$.

One may try to add a free lapse in the interior, $f_{-}=k^{2}$ with $k\neq1$, to
control the rate of interior clocks~\cite{BobrickMartire2021,Fuchs2024}. However,
the metric~\eqref{eq:bulk} then has Ricci scalar $R=2(1-k^{2})/r^{2}$ and
describes a global monopole~\cite{BarriolaVilenkin1989}, with a solid-angle deficit
and a curvature singularity at $r=0$. A flat payload region requires $k=1$. This
does not remove the clock effect but fixes it: for a static shell
$\dd t_{-}/\dd t_{+}=\sqrt{1-2M/a_{0}}$, so interior clocks run slow by a factor
set by the compactness.

Boosting the configuration with velocity $v$ gives a flat region moving at speed
$v$ with respect to the asymptotic frame, with no proper acceleration or tidal
stress inside. This is the content of a subluminal warp
drive~\cite{BobrickMartire2021}. The boost, however, is a diffeomorphism of an
asymptotically flat solution, and the moving shell is the static one seen from a
different frame. Fuchs \emph{et al.}~\cite{Fuchs2024} check that the shift vector
of their solution cannot be removed by a coordinate transformation; by this
criterion the present shell is the degenerate member of the class. The positive
side of this is that all the invariant quantities used below, namely the eigenvalues of
the surface stress tensor, the effective potential and the stability
threshold, are independent of $v$. In Kerr--Schild form,
$g_{\mu\nu}=\eta_{\mu\nu}+(2M/r)l_{\mu}l_{\nu}$, a Lorentz transformation preserves
$\eta_{\mu\nu}$ and maps the null vector $l$ to another null vector, so the boosted
exterior is Schwarzschild with the same $M$. The energy measured in the asymptotic
frame is $\gamma M$, which is a property of the observer and not of the
shell~\cite{SchusterSantiagoVisser2023}. The acceleration phase lies outside this
construction. All the nontrivial physics is in the shell, which we study in its
rest frame.

%% ===================================================================
\section{Surface stresses and energy conditions}
\label{sec:junction}
%% ===================================================================

Let $r=a(\tau)$ be the shell trajectory, with $\tau$ its proper time, so that the
induced metric is $\dd s^{2}_{\Sigma}=-\dd\tau^{2}+a^{2}\dd\Omega^{2}$. The
nonvanishing components of the extrinsic curvature are
\begin{equation}
  K^{\tau}{}_{\tau\,\pm}=\frac{\ddot a+\half f'_{\pm}}{\sqrt{f_{\pm}+\dot a^{2}}},
  \qquad
  K^{\theta}{}_{\theta\,\pm}=K^{\varphi}{}_{\varphi\,\pm}
   =\frac{\sqrt{f_{\pm}+\dot a^{2}}}{a}.
  \label{eq:K}
\end{equation}
We take the wall to be a perfect fluid, so that
$S^{i}{}_{j}=\mathrm{diag}(-\sigma,p,p)$ in its rest frame. The Lanczos equation
$S^{i}{}_{j}=-\frac{1}{8\pi}(\jump{K^{i}{}_{j}}-\delta^{i}{}_{j}\jump{K})$ gives
\begin{align}
  \sigma &= -\frac{1}{4\pi a}
    \left[\sqrt{f_{+}+\dot a^{2}}-\sqrt{f_{-}+\dot a^{2}}\right],
  \label{eq:sigma}\\
  p &= \frac{1}{8\pi}\left[
      \frac{2\ddot a+f'_{+}}{2\sqrt{f_{+}+\dot a^{2}}}
    + \frac{\sqrt{f_{+}+\dot a^{2}}}{a}\right.\nonumber\\
    &\qquad\quad\left.
    - \frac{2\ddot a+f'_{-}}{2\sqrt{f_{-}+\dot a^{2}}}
    - \frac{\sqrt{f_{-}+\dot a^{2}}}{a}\right].
  \label{eq:p}
\end{align}
It is convenient to introduce
\begin{equation}
  s \equiv \sqrt{1-\frac{2M}{a_{0}}}\in(0,1),
  \qquad \frac{a_{0}}{M}=\frac{2}{1-s^{2}},
  \label{eq:sdef}
\end{equation}
so that $s\to0$ is the horizon limit and $s\to1$ the weak-field limit. For a
static shell, $\dot a=\ddot a=0$, and in units $M=1$,
\begin{equation}
  \sigma_{0}=\frac{(1-s)^{2}(1+s)}{8\pi},
  \qquad
  p_{0}=\frac{(1-s)^{3}(1+s)}{32\pi s}.
  \label{eq:sigp0}
\end{equation}
Both are positive for $0<s<1$. For $s\to1$ we recover a Newtonian dust shell,
$\sigma_{0}\to M/4\pi a_{0}^{2}$ and $p_{0}=O(M^{2}/a_{0}^{3})$. The rest mass of
the shell, $m_{s}\equiv4\pi a_{0}^{2}\sigma_{0}=a_{0}(1-s)$, satisfies
\begin{equation}
  M = m_{s}-\frac{m_{s}^{2}}{2a_{0}},
  \label{eq:fundrel}
\end{equation}
which states exactly that the gravitational mass is the rest mass minus the
binding energy.

The conservation law $\nabla_{i}S^{i}{}_{j}=0$ reads
\begin{equation}
  \dot\sigma+\frac{2\dot a}{a}(\sigma+p)=0
  \quad\Longleftrightarrow\quad m_{s}'(a)=-8\pi a\,p(a),
  \label{eq:cons}
\end{equation}
i.e.\ the first law $\dd m_{s}=-p\,\dd A$ with $A=4\pi a^{2}$, at fixed entropy.
For a barotropic wall $p=p(\sigma)$ with squared sound speed
$\beta^{2}\equiv\dd p/\dd\sigma$ we then have
\begin{equation}
  m_{s}''(a)=-8\pi p+16\pi\beta^{2}(\sigma+p).
  \label{eq:cons2}
\end{equation}
We assume $0\le\beta^{2}\le1$, i.e.\ that perturbations propagate causally in the
wall, as in Refs.~\cite{BradyLoukoPoisson1991,IshakLake2002,LoboCrawford2005}.
This is an extra physical requirement, not implied by the energy conditions.

\begin{figure}[t]
\includegraphics[width=\columnwidth]{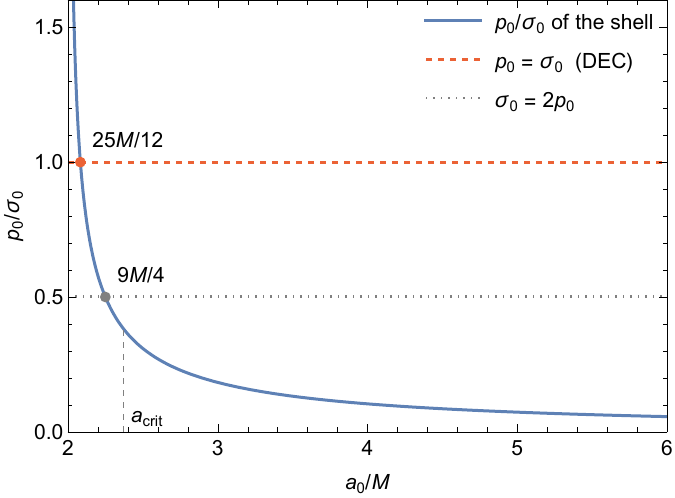}
\caption{Equation of state of the static shell, $p_{0}/\sigma_{0}=(1-s)/4s$. The
null, weak and strong energy conditions hold for all $a_{0}>2M$; the dominant one
fails for $a_{0}<25M/12$. The dotted line $\sigma_{0}=2p_{0}$ is crossed at
$a_{0}=9M/4$. Every radially stable shell ($a_{0}>a_{\rm crit}$) lies below both
lines.}
\label{fig:eos}
\end{figure}

Since $S^{i}{}_{j}$ is diagonal with real eigenvalues $\{-\sigma,p,p\}$, it is of
Hawking--Ellis type~I and each energy condition reduces to an inequality on
$\sigma$ and $p$~\cite{HawkingEllis1973,MartinMorunoVisser2017}. From
Eq.~\eqref{eq:sigp0},
\begin{align}
  \sigma_{0}+p_{0}  &= \frac{(1-s)^{2}(1+s)(1+3s)}{32\pi s}, \label{eq:nec}\\
  \sigma_{0}+2p_{0} &= \frac{(1-s)^{2}(1+s)^{2}}{16\pi s},   \label{eq:sec}\\
  \sigma_{0}-p_{0}  &= \frac{(1-s)^{2}(1+s)(5s-1)}{32\pi s}. \label{eq:dec}
\end{align}
The null, weak and strong energy conditions therefore hold for every $a_{0}>2M$.
The dominant energy condition, $\sigma_{0}\ge p_{0}$, holds if and only if
$s\ge1/5$, that is
\begin{equation}
  a_{0}\ \ge\ \frac{25}{12}\,M \simeq 2.083\,M,
  \label{eq:decbound}
\end{equation}
in agreement with the minimum radius found in Ref.~\cite{FHK1990}. The ratio
$p_{0}/\sigma_{0}=(1-s)/4s$ is shown in Fig.~\ref{fig:eos}. The diagonal form of
$S^{i}{}_{j}$ follows from the perfect-fluid assumption: the flat-payload
geometry admits a type~I source, in contrast with the Alcubierre, Nat\'ario and
Van Den Broeck walls, but a wall with heat flux or anisotropic stress need not be
of this type.

%% ===================================================================
\section{Radial stability}
\label{sec:dynamics}
%% ===================================================================

Equation~\eqref{eq:sigma} can be rewritten as $\dot a^{2}+V(a)=0$ with
\begin{align}
  V(a)&=f_{-}-\left[\frac{m_{s}}{2a}+\frac{a(f_{-}-f_{+})}{2m_{s}}\right]^{2}
  \nonumber\\
      &=1-\left[\frac{m_{s}(a)}{2a}+\frac{M}{m_{s}(a)}\right]^{2},
  \label{eq:V}
\end{align}
where we used $f_{-}-f_{+}=2M/a$. With Eqs.~\eqref{eq:sigp0} and \eqref{eq:cons}
one finds $V(a_{0})=V'(a_{0})=0$ identically, and the shell is stable if
$V''(a_{0})>0$~\cite{PoissonVisser1995,IshakLake2002}. Using
Eqs.~\eqref{eq:cons}--\eqref{eq:cons2}, $V''(a_{0})$ is linear in $\beta^{2}$,
\begin{equation}
  V''(a_{0})=\mathcal{A}(s)+\mathcal{B}(s)\,\beta^{2},
  \label{eq:Vpp}
\end{equation}
with (units $M=1$)
\begin{align}
  \mathcal{A}(s)&=-\frac{(1-s)^{3}(1+s)^{2}\left(1+2s+3s^{2}\right)}{8s^{2}},
  \label{eq:Acoef}\\
  \mathcal{B}(s)&=\frac{(1-s)^{2}(1+s)^{2}(1+3s)}{2}
   =\frac{8s\,(\sigma_{0}+p_{0})}{a_{0}^{2}\,\sigma_{0}}.
  \label{eq:Bcoef}
\end{align}
The second form of $\mathcal{B}$, derived in Appendix~\ref{app:general}, shows
that $\mathcal{B}>0$ exactly when the null energy condition holds. Stability is
therefore always a lower bound on the stiffness of the wall,
$\beta^{2}>\beta^{2}_{\rm crit}$, with
\begin{equation}
  \beta^{2}_{\rm crit}(s)
   =\frac{1+s+s^{2}-3s^{3}}{4s^{2}\left(1+3s\right)}.
  \label{eq:b2crit}
\end{equation}
This diverges at the horizon, $s\to0$, and goes to zero in the weak-field limit,
$s\to1$, where a very soft wall is enough.

\subsection{The threshold as the slope of the static sequence}
\label{sec:sequence}

Equation~\eqref{eq:b2crit} has a simple origin. Consider the one-parameter
sequence of static solutions labelled by $a_{0}$, at fixed $M$, and its rest mass
\begin{equation}
  \mathcal{M}_{s}(a)\equiv4\pi a^{2}\sigma_{0}(a)
  =a\left[\sqrt{f_{-}(a)}-\sqrt{f_{+}(a)}\right].
  \label{eq:Mseq}
\end{equation}
Using the static form of Eqs.~\eqref{eq:sigma}--\eqref{eq:p} one finds, for
arbitrary $f_{\pm}$,
\begin{equation}
  \mathcal{M}_{s}'(a)=-8\pi a\,p_{0}(a),
  \qquad
  \frac{\dd\sigma_{0}}{\dd a}=-\frac{2}{a}\left(\sigma_{0}+p_{0}\right),
  \label{eq:seqODE}
\end{equation}
which is the same equation~\eqref{eq:cons} obeyed by the dynamical $m_{s}(a)$.
Hence $m_{s}$ and $\mathcal{M}_{s}$ coincide at $a_{0}$ together with their first
derivatives, whatever the wall material, and since $V$ vanishes along the static
sequence this is why $V(a_{0})=V'(a_{0})=0$. They also agree at second order, so
that $V''(a_{0})=0$, precisely when $\beta^{2}\,\dd\sigma_{0}/\dd a=\dd p_{0}/\dd a$.
Therefore, for any pair of bulk metrics,
\begin{equation}
  \beta^{2}_{\rm crit}
   =\left.\frac{\dd p_{0}}{\dd\sigma_{0}}\right|_{\rm static\ sequence}.
  \label{eq:slope}
\end{equation}
Marginal stability corresponds to the existence of a neighbouring equilibrium,
and a shell is stable when its wall is stiffer than the local slope of the
sequence in the $(\sigma_{0},p_{0})$ plane. This is the thin-shell version of the
turning-point criterion of stellar structure. Differentiating Eq.~\eqref{eq:sigp0},
$\dd\sigma_{0}/\dd s=(3s^{2}-2s-1)/8\pi$ and
$\dd p_{0}/\dd s=(4s^{3}-3s^{4}-1)/32\pi s^{2}$, whose ratio is
Eq.~\eqref{eq:b2crit}.

\subsection{The causal threshold}

\begin{figure}[t]
\includegraphics[width=\columnwidth]{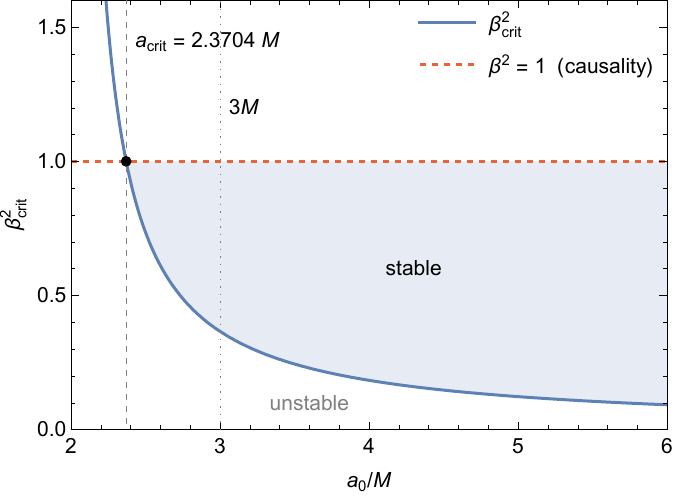}
\caption{Critical squared sound speed of the wall, Eq.~\eqref{eq:b2crit}, which is
also the slope $\dd p_{0}/\dd\sigma_{0}$ of the static sequence. Shells above the
curve are radially stable. The shaded region is the part allowed by causality,
$\beta^{2}\le1$, which begins at $a_{\rm crit}=2.3704\,M$. Note that
$a_{\rm crit}$ lies inside the photon sphere $3M$ of the exterior.}
\label{fig:b2crit}
\end{figure}

Setting $\beta^{2}_{\rm crit}=1$ in Eq.~\eqref{eq:b2crit} gives
$1+s+s^{2}-3s^{3}=4s^{2}+12s^{3}$, so that a shell made of causal matter is
radially stable if and only if $s>s_{\ast}$, where $s_{\ast}$ is the only real
root of
\begin{equation}
  15s^{3}+3s^{2}-s-1=0 .
  \label{eq:cubic}
\end{equation}
By Cardano's formula,
\begin{align}
  s_{\ast}&=\frac{1}{15}\left[\sqrt[3]{104+10\sqrt{106}}
   +\sqrt[3]{104-10\sqrt{106}}-1\right]\nonumber\\
   &=0.3952950,
  \label{eq:radical}
\end{align}
which corresponds to
\begin{equation}
  \frac{a_{\rm crit}}{M}=\frac{2}{1-s_{\ast}^{2}}=2.3703931 .
  \label{eq:acrit}
\end{equation}
No causal wall can stabilise a shell inside $a_{\rm crit}$, i.e.\ for compactness
$2M/a_{0}\gtrsim0.844$. Since $a_{\rm crit}>25M/12$, every radially stable shell
also satisfies the dominant energy condition. Outside the threshold the required
stiffness drops quickly: $\beta^{2}_{\rm crit}=0.366$, $0.123$, $0.0465$ and
$0.0162$ at $a_{0}=3M$, $5M$, $10M$ and $25M$ (Fig.~\ref{fig:b2crit}).

Brady, Louko and Poisson observed that the surface density of a stable shell
always exceeds twice its pressure~\cite{BradyLoukoPoisson1991}. Here
\begin{equation}
  \sigma_{0}-2p_{0}=\frac{(1-s)^{2}(1+s)(3s-1)}{16\pi s},
  \label{eq:sig2p}
\end{equation}
which is positive for $s>1/3$, i.e.\ $a_{0}>9M/4$. Since $a_{\rm crit}>9M/4$, the
observation holds here as well. The value $9M/4$ coincides with the Buchdahl
radius, which is known to appear also as a dynamical and thermodynamic radius for
thin shells~\cite{AndreLemos2021}. A thin shell is not a Buchdahl fluid ball,
however, and in the variables of Sec.~\ref{sec:interp} the condition
$\sigma_{0}=2p_{0}$ is simply $u=1/3$.

%% ===================================================================
\section{A black hole inside the shell}
\label{sec:interp}
%% ===================================================================

The threshold~\eqref{eq:acrit} lies inside the exterior photon sphere $3M$, while
Ref.~\cite{BradyLoukoPoisson1991} states that a stable shell always lies outside
the circular photon orbit. There is no contradiction: in that work the orbit is
that of the interior black hole, $r=3m$, which is absent for $m=0$. To see this
explicitly we replace the flat interior by $f_{-}=1-2m/r$ with $0\le m<M$, which
gives the family of Ref.~\cite{BradyLoukoPoisson1991}. We define
\begin{equation}
  x\equiv\sqrt{1-\frac{2m}{a}},\qquad y\equiv\sqrt{1-\frac{2M}{a}},
  \qquad u\equiv xy,
  \label{eq:xydef}
\end{equation}
so that $0<y<x\le1$ and $u=s$ for $m=0$. The static stresses become
\begin{equation}
  \sigma_{0}=\frac{x-y}{4\pi a},\qquad
  p_{0}=\frac{(x-y)(1-u)}{16\pi a\,u},\qquad
  \frac{p_{0}}{\sigma_{0}}=\frac{1-u}{4u},
  \label{eq:eosgen}
\end{equation}
so the equation of state depends on $u$ only. The null, weak and strong energy
conditions hold for every $m<M$, the dominant one holds for $u\ge1/5$, and
$\sigma_{0}\ge2p_{0}$ for $u\ge1/3$. Using Eq.~\eqref{eq:slope}
(Appendix~\ref{app:general}) the critical sound speed is
\begin{equation}
  \beta^{2}_{\rm crit}=\frac{x^{2}+y^{2}+u-3u^{3}}{4u^{2}\left(1+3u\right)},
  \label{eq:b2critgen}
\end{equation}
and the causal threshold $\beta^{2}_{\rm crit}=1$ is the curve
\begin{equation}
  \Phi(x,y)\equiv x^{2}+y^{2}+xy-4x^{2}y^{2}-15x^{3}y^{3}=0 .
  \label{eq:Phi}
\end{equation}
For $m=0$ ($x=1$, $y=s$) this gives back the cubic~\eqref{eq:cubic}. For $m\to M$
($x=y$, $u=x^{2}$) it becomes $u(3u-1)(5u+3)=0$, so the threshold ends at $u=1/3$,
that is at $a=3M$, where the shell becomes massless.

\medskip
\noindent\textbf{Proposition.} \emph{For $0\le m<M$ the threshold satisfies
$3m<a_{\rm crit}(m)<3M$ and increases strictly with $m$, from $2.3703931\,M$ at
$m=0$ to $3M$ as $m\to M$. Along the threshold
$1/3\le u\le s_{\ast}$.}
\medskip

\noindent\emph{Proof.} Note that $a>3m$ is equivalent to $x^{2}>1/3$, and $a<3M$
to $y^{2}<1/3$.
(i) Since $x^{2}+y^{2}\ge2u$, we have
$\Phi\ge3u-4u^{2}-15u^{3}=-15u(u-\tfrac13)(u+\tfrac35)$, which is positive for
$u<1/3$. Hence $\Phi=0$ requires $u\ge1/3$, and $x^{2}\ge xy=u\ge1/3$, with
equality only for $x=y$, i.e.\ $m=M$.
(ii) If $y^{2}\ge1/3$, then $u\ge1/3$ and for $x\ge y$
$\partial\Phi/\partial x=x(2-8y^{2})+y(1-45u^{2})<0$, while
$\Phi(y,y)=-15y^{2}(y^{2}-\tfrac13)(y^{2}+\tfrac35)\le0$. Thus $\Phi<0$ whenever
$y^{2}\ge1/3$ and $x>y$, and the threshold must have $y^{2}<1/3$.
(iii) On $\Phi=0$ one has $x^{2}y^{2}=u^{2}$ and
$x^{2}+y^{2}=\Psi(u)\equiv15u^{3}+4u^{2}-u$, so $P\equiv x^{2}$ and
$Q\equiv y^{2}$ are the roots of $T^{2}-\Psi(u)\,T+u^{2}=0$. They are real for
$u\ge1/3$, and $P\le1$ for $u\le s_{\ast}$, so $u\in[1/3,s_{\ast}]$ parametrises
the curve once. Differentiating,
$\dd P/\dd u=(P\Psi'-2u)/(P-Q)$ and $\dd Q/\dd u=(2u-Q\Psi')/(P-Q)$, with
$\Psi'=45u^{2}+8u-1\ge20/3$ on this interval. Then $P\Psi'-2u\ge20/9-2s_{\ast}>0$ because $P\ge1/3$, and
$Q\Psi'-2u\ge u(45u^{3}+8u^{2}-u-2)>0$ because $Q=u^{2}/P\ge u^{2}$, the cubic
being $2/9$ at $u=1/3$ and increasing. Hence $\dd P/\dd u>0$ and $\dd Q/\dd u<0$. Since $a_{\rm crit}/M=2/(1-Q)$ and
$m/M=(1-P)/(1-Q)$, both decrease as $u$ grows, so $a_{\rm crit}$ increases with
$m$.~$\square$
\medskip

\begin{figure}[t]
\includegraphics[width=\columnwidth]{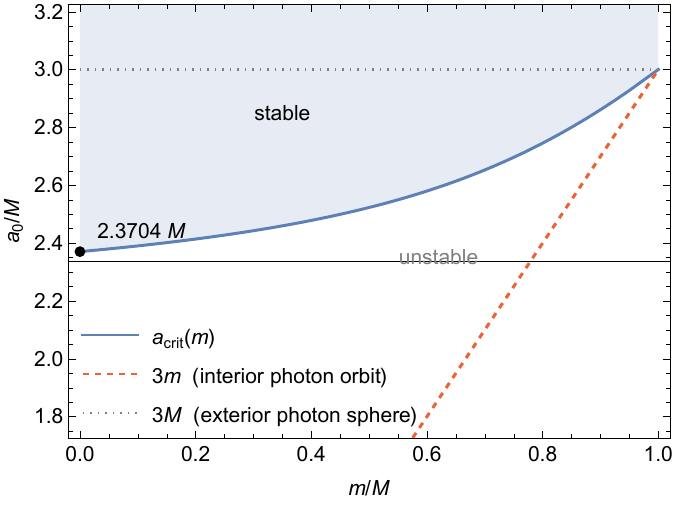}
\caption{Radial stability threshold for causal wall material as a function of the
interior mass $m$, from the flat interior ($m=0$) to the massless-shell limit
$m\to M$. Shells above the curve are stable. The threshold rises monotonically to
the exterior photon sphere $3M$ and stays above the interior photon orbit $3m$
(dashed).}
\label{fig:interp}
\end{figure}

Table~\ref{tab:interp} and Fig.~\ref{fig:interp} show the result: $u$ changes by
less than $19\%$ along the whole curve, while $a_{\rm crit}-3m$ decreases from
$2.37M$ to zero. Two consequences follow. First, since
$u^{2}=(1-2m/a)(1-2M/a)$ grows with $a$, every radially stable shell made of
causal matter has $u>1/3$, and therefore $\sigma_{0}>2p_{0}$ and satisfies all
the energy conditions, for any interior mass. This turns the observation of
Ref.~\cite{BradyLoukoPoisson1991} into a statement valid for the whole family.
Second, the flat interior gives the lowest threshold of the family. The members
with $m>0$ contain a black hole and are not warp drives, but the comparison shows
that the warp configuration is the most favourable case for radial stability.

\begin{table}[t]
\caption{Radial stability threshold for causal wall material as a function of the
interior mass, and the value of $u=\sqrt{f_{-}f_{+}}$ at the threshold.}
\begin{ruledtabular}
\begin{tabular}{ccccc}
$m/M$ & $a_{\rm crit}/M$ & $3m/M$ & $(a_{\rm crit}-3m)/M$ & $u$ \\ \hline
0      & 2.3703931 & 0      & 2.37039  & 0.3952950 \\
0.1    & 2.3902397 & 0.3    & 2.09024  & 0.3867854 \\
0.3    & 2.4427835 & 0.9    & 1.54278  & 0.3697838 \\
0.5    & 2.5231027 & 1.5    & 1.02310  & 0.3537718 \\
0.7    & 2.6528960 & 2.1    & 0.55290  & 0.3409253 \\
0.9    & 2.8611669 & 2.7    & 0.16117  & 0.3341122 \\
0.99   & 2.9851077 & 2.97   & 0.01511  & 0.3333405 \\
0.9999 & 2.9998500 & 2.9997 & 0.00015  & 0.3333333 \\
\end{tabular}
\end{ruledtabular}
\label{tab:interp}
\end{table}

%% ===================================================================
\section{Discussion}
\label{sec:discussion}
%% ===================================================================

We have obtained the radial stability of the simplest subluminal warp shell in analytical form. The shell needs no exotic matter, satisfies all the energy
conditions outside $25M/12$, and can be stabilised by causal wall material only
outside $a_{\rm crit}=2.3703931\,M$, the lowest value in the family of shells
around a black hole. The critical sound speed is the slope of the static
sequence, a result that holds for any pair of bulk metrics.

It is worth noting where this threshold matters. Since $\beta^{2}_{\rm crit}$
falls below $0.02$ already at $a_{0}=25M$, a bubble of macroscopic size and modest
mass, $a_{0}\gg M$, is radially stable for almost any wall with nonzero stiffness.
The threshold constrains only objects close to their own horizon. Its natural
setting is therefore that of ultracompact objects and black-hole
mimickers~\cite{Pani2009,YangBongaPan2023,PSP2026,Efstratiou2026}, where the same
shells are under active study.

Radial stability is not enough for viability. Pitre, Schneider and
Poisson~\cite{PSP2026} found an exponentially growing even-parity mode for this
configuration for all sampled values of the compactness, the adiabatic index and
the multipole number $\ell\ge2$, extending the eikonal result of
Ref.~\cite{YangBongaPan2023}. Their survey is numerical, but no stable member has
been found. Since the boost does not change the shell, setting the bubble in
motion cannot remove the instability, and since the flat interior is the best
case of the family, the problem cannot be cured by changing the interior mass.
The threshold~\eqref{eq:acrit} then marks the boundary of the region in which a
stabilising mechanism would have to work.

A few directions remain open. The analysis of Ref.~\cite{PSP2026} assumes a perfect
fluid, and a wall with shear rigidity, which resists the transverse deformations
that drive the $\ell\ge2$ modes, is not covered by it. The same work notes a
tension with the thin-shell gravastar of Ref.~\cite{Pani2009}, which has a de
Sitter interior and showed no unstable modes. The two models differ both in the
interior and in the wall, since the gravastar shell has a fixed equation of state
with no free $\beta^{2}$. Our formalism depends on the bulk only through
$f_{\pm}$, and Eq.~\eqref{eq:slope} holds for any $f_{\pm}$, so it applies directly
to a de Sitter interior. Finally, our treatment is isentropic. In the thermodynamic
approach of Ref.~\cite{AndreLemosQuinta2019} a shell with a Hawking-like
temperature has positive heat capacity only between its gravitational radius and
the photon sphere, which is also where $a_{\rm crit}(m)$ ends. It would be
interesting to see whether the turning-point picture of Eq.~\eqref{eq:slope}
carries over when the static sequence is followed at fixed entropy.

\begin{acknowledgments}
AI acknowledges financial support from SECIHTI through the National Postgraduate
Scholarship Program (CVU 2222058).
\end{acknowledgments}

\appendix

%% ===================================================================
\section{The general-\texorpdfstring{$m$}{m} criterion}
\label{app:general}
%% ===================================================================

Let $f_{-}=1-2m/r$, $f_{+}=1-2M/r$ and $x$, $y$, $u$ as in Eq.~\eqref{eq:xydef}.
Writing $F\equiv m_{s}/2a+(M-m)/m_{s}$, so that $V=f_{-}-F^{2}$, the term in
$V''=f_{-}''-2F'^{2}-2FF''$ that contains $m_{s}''$ is
$-2F\,m_{s}''\left[1/2a-(M-m)/m_{s}^{2}\right]$. On the static sequence
$m_{s}=a(x-y)$ and $M-m=\half a(x^{2}-y^{2})$, so that $F=x$ and the bracket equals
$-y/[a(x-y)]$. With Eq.~\eqref{eq:cons2} and $x-y=4\pi a\sigma_{0}$ this gives
\begin{equation}
  \mathcal{B}=\frac{2u}{a(x-y)}\,16\pi(\sigma_{0}+p_{0})
             =\frac{8u\,(\sigma_{0}+p_{0})}{a^{2}\sigma_{0}},
  \label{eq:Bgen}
\end{equation}
which reduces to Eq.~\eqref{eq:Bcoef} for $m=0$.

For $\beta^{2}_{\rm crit}$ we use Eq.~\eqref{eq:slope}. Writing
$p_{0}=\sigma_{0}\,g(u)$ with $g=(1-u)/4u$,
\begin{equation}
  \beta^{2}_{\rm crit}
   =g+\sigma_{0}\,g'(u)\,\frac{u'}{\sigma_{0}'},
\end{equation}
where a prime denotes $\dd/\dd a$ on $\sigma_{0}$ and $u$. From
Eq.~\eqref{eq:seqODE}, $\sigma_{0}/\sigma_{0}'=-2au/(1+3u)$, while $u^{2}=f_{-}f_{+}$
gives $u'=(x^{2}+y^{2}-2u^{2})/2au$. With $g'=-1/4u^{2}$,
\begin{equation}
  \beta^{2}_{\rm crit}=\frac{1-u}{4u}+\frac{x^{2}+y^{2}-2u^{2}}{4u^{2}(1+3u)}
   =\frac{x^{2}+y^{2}+u-3u^{3}}{4u^{2}(1+3u)},
\end{equation}
which is Eq.~\eqref{eq:b2critgen}. We have checked symbolically that this agrees
with $-\mathcal{A}/\mathcal{B}$ computed directly from Eq.~\eqref{eq:V}; the
accompanying Mathematica notebook reproduces all the equations and figures, accessible by reasonable request.

\bibliographystyle{apsrev4-2}
\bibliography{refs}

\end{document}